\documentclass{aa}  

\usepackage{makecell}
\usepackage{graphicx}
\usepackage{natbib}
\usepackage[utf8]{inputenc}
\usepackage{natbib}
\usepackage[mathscr]{euscript}
\usepackage{lscape}

\usepackage{booktabs,caption,fixltx2e}
\usepackage[flushleft]{threeparttable}
\begin{document}

   \title{The chemical connection between damped Lyman-$\alpha$ systems and Local Group dwarf galaxies}
        \titlerunning{The chemical connection between DLA systems and dwarf galaxies}
   
      \author{\'{A}. Sk\'{u}lad\'{o}ttir
                \inst{1}
        \and            
         S. Salvadori\inst{2,3,4} 
        \and
            M. Pettini \inst{5}   
         \and     
        E. Tolstoy\inst{6}
        \and
         V. Hill \inst{7}
          }
          
   \institute{
                        Max-Planck-Institut f$\ddot{\text{u}}$r Astronomie, K$\ddot{\text{o}}$nigstuhl 17, D-69117 Heidelberg, Germany\\
                        \email{skuladottir@mpia.de}
                \and
                Dipartimento di Fisica e Astronomia, Universita’ di Firenze, Via G. Sansone 1, Sesto Fiorentino, Italy
                \and 
                INAF/Osservatorio Astrofisico di Arcetri, Largo E. Fermi 5, Firenze, Italy
                \and
                GEPI, Observatoire de Paris, PSL Research University, CNRS, Place Jule Janssen 92190, Meudon, France
                     \and
                Institute of Astronomy, Madingley Road, Cambridge CB3 0HA, England.   
                                 \and
              Kapteyn Astronomical Institute, University of Groningen, PO Box 800, 9700AV Groningen, the Netherlands
              \and   
             Laboratoire Lagrange, Universit\'{e} de Nice Sophia Antipolis, CNRS, Observatoire de la C\^{o}te d’Azur, CS34229, 06304 Nice Cedex 
                        4, France
                }


\abstract{

Abundances of the volatile elements S and Zn have now been measured in around 80 individual stars in the Sculptor dwarf spheroidal galaxy, covering the metallicity range $-2.4\leq\text{[Fe/H]}\leq-0.9$. These two elements are of particular interest as they are not depleted onto dust in gas, and their ratio, [S/Zn], has thus commonly been used as a proxy for [$\alpha$/Fe] in Damped Lyman-$\alpha$ systems (DLAs). The S abundances in Sculptor are similar to other $\alpha$-elements in this galaxy, consistent with S being mainly created in core-collapse supernovae, but also having some contribution from type Ia supernovae. However, our results show that Zn and Fe do not trace all the same nucleosynthetic production channels. In particular, (contrary to Fe) Zn is not significantly produced by type Ia supernovae. Thus, [S/Zn] cannot be reliably used as a proxy for [$\alpha$/Fe]. We propose [O/S] as a function of [S/H] as a possible alternative. At higher metallicities, the values of [S/Zn] measured in DLAs are inconsistent with those in local dwarf galaxies, and are more compatible with the Milky Way disk. Low-metallicity  DLAs are, however, consistent with the most metal-poor stars in Local Group dwarf spheroidal galaxies. Assuming that the dust depletions of S and Zn are negligible, our comparison indicates that the star formation histories of DLAs are on average different from both the Milky Way and the Sculptor dwarf spheroidal galaxy.
}               
               
                \maketitle

%
\section{Introduction}

Connecting the near and the far has always been one of astronomy's main challenges. When it comes to star formation, in the Local Group we are able to resolve individual stars with spectroscopy and/or photometry and thus study the time-dependent details of stellar populations in precise detail. However, as we go further in distance and back in time, such studies quickly become too time consuming or impossible with current astronomical instrumentation. Instead, we are limited to analyzing the integrated light of a whole galaxy to gather information on the history of its stellar populations. Even then, only the brightest galaxies are visible, while dwarf galaxies are generally expected to become too faint to observe with both current and planned instrumentation (e.g., \citealt{Pettini2014,Wise2014,Boylan-Kolchin2015}). Luckily, intervening absorption systems can be observed in the light of bright background sources (usually quasars (QSO) or gamma-ray bursts (GRB)), up to very high redshifts, $z\approx6$ and beyond (e.g., \citealt{Rafelski2014,Becker2015}).

The heavy end of the spectrum of such systems are the damped Lyman-$\alpha$ systems (DLAs), defined to have column densities of neutral hydrogen $N(\text{H I}) \geq 2 \times 10^{20}$~cm$^{-2}$. At these high column densities, the gas is self-shielding from ionizing radiation, and thus the hydrogen is mostly neutral, while in other classes of QSO absorption systems a significant fraction of the gas is ionized (e.g., \citealt{Bechtold2003}). Furthermore, it has been shown that DLAs contain the majority of neutral hydrogen in the universe at all redshifts (e.g., \citealt{Wolfe2005,Neeleman2016,Sanchez-Ramirez2016}), and since stars form from cold neutral clouds and not warm ionized gas (e.g., \citealt{Wolfire2003}), DLAs are thought to play an important role in the overall star formation history of the universe (e.g., \citealt{Wolfe2005}).

Identifying DLA systems is straightforward even in low-resolution spectra (e.g., \citealt{Mas-Ribas2017}), however, the true nature of the underlying galaxies has proved difficult to constrain. Imaging at low redshift, $z<1$, has revealed both low-luminosity and low-surface-brightness galaxies, as well as brighter spiral galaxies (e.g., \citealt{Krogager2017} and references therein). Thus, DLAs seem to arise in a wide range of galaxies, the common requirement being a high surface density of neutral gas \citep{Pontzen2008,Fumagalli2015,Bird2015}.

With increasing redshift, there is a moderate effect of decreasing mean metallicity in the observed sample of DLA systems, going from $\text{<[M/H]>}\gtrsim-1$ at $z=1$, to $\text{<[M/H]>}\approx-2$ at $z=5$ (\citealt{Rafelski2012,Rafelski2014}). However, at any given redshift there is also a large scatter, exceeding the measurement errors. In the redshift range where DLA measurements are most abundant, $2\leq z \leq 4$, this scatter covers up to 2 dex in gas metallicity (\citealt{Rafelski2012,Rafelski2014,Jorgenson2013}), thus indicating that DLAs are not a uniform population of galaxies evolving in a homogeneous manner. 

In particular, a link has been proposed between the metal-poor DLA population and Local Group dwarf galaxies (e.g., \citealt{SalvadoriFerrara2012,Cooke2015,Berg2015}). For nearby galaxies there is a well known mass-metallicity relation: the more massive a galaxy, the higher its metallicity. This holds true whether the metallicities are measured in gas phase (\citealt{Skillman1989,Lee2006}) or stars \citep{Kirby2013}. A comparable relation has been shown to exist in DLAs, where the least massive systems are those with the lowest metallicities \citep{Peroux2012,Christensen2014,Krogager2017}.

In principle, abundance measurements in DLAs are relatively simple, as most of the gas is neutral and atoms are in a single dominant ionization stage and the ionization corrections are therefore negligible. However, one of the major challenges comes from the fact that a fraction of the metals in the interstellar medium (ISM) is expected to be depleted onto dust (\citealt{Savage1996,Jenkins2009,Jenkins2017,Cia2016}). Therefore, the abundance measurements of refractory elements, such as Fe and Cr, in DLAs underestimate their true values, and uncertain dust corrections need to be applied. However, the more volatile elements, in particular N, O, S, and Zn, are normally not significantly depleted onto dust at the metallicities and H I column densities of most DLAs (e.g., \citealt{Tchernyshyov2015,Cia2016,Jenkins2017}), and are therefore especially important tracers of the chemical properties of DLA systems, and absorption systems in general.

Two main contributors of metals in galaxies are core-collapse Supernovae (SN) (such as type II), that explode on short time scales ($\sim10^{6-7}$~yr) and enrich the environment primarily with $\alpha$-elements, such as O, Mg, Si, and S (\citealt{Nomoto2013}). On the other hand,  type Ia SN release mainly Fe-peak elements and typically explode around 1-2~Gyr after the onset of star formation (e.g., \citealt{MatteucciGreggio1986,WyseGilmore1988,deBoer2012}). The abundance ratios of [$\alpha$/Fe] thus provide information about the relative fractions of type Ia and II SN in a galaxy, as well as the time scales of star formation (e.g., \citealt{Wheeler1989}).

     \begin{figure*}
   \centering
   \includegraphics[width=\hsize-1cm]{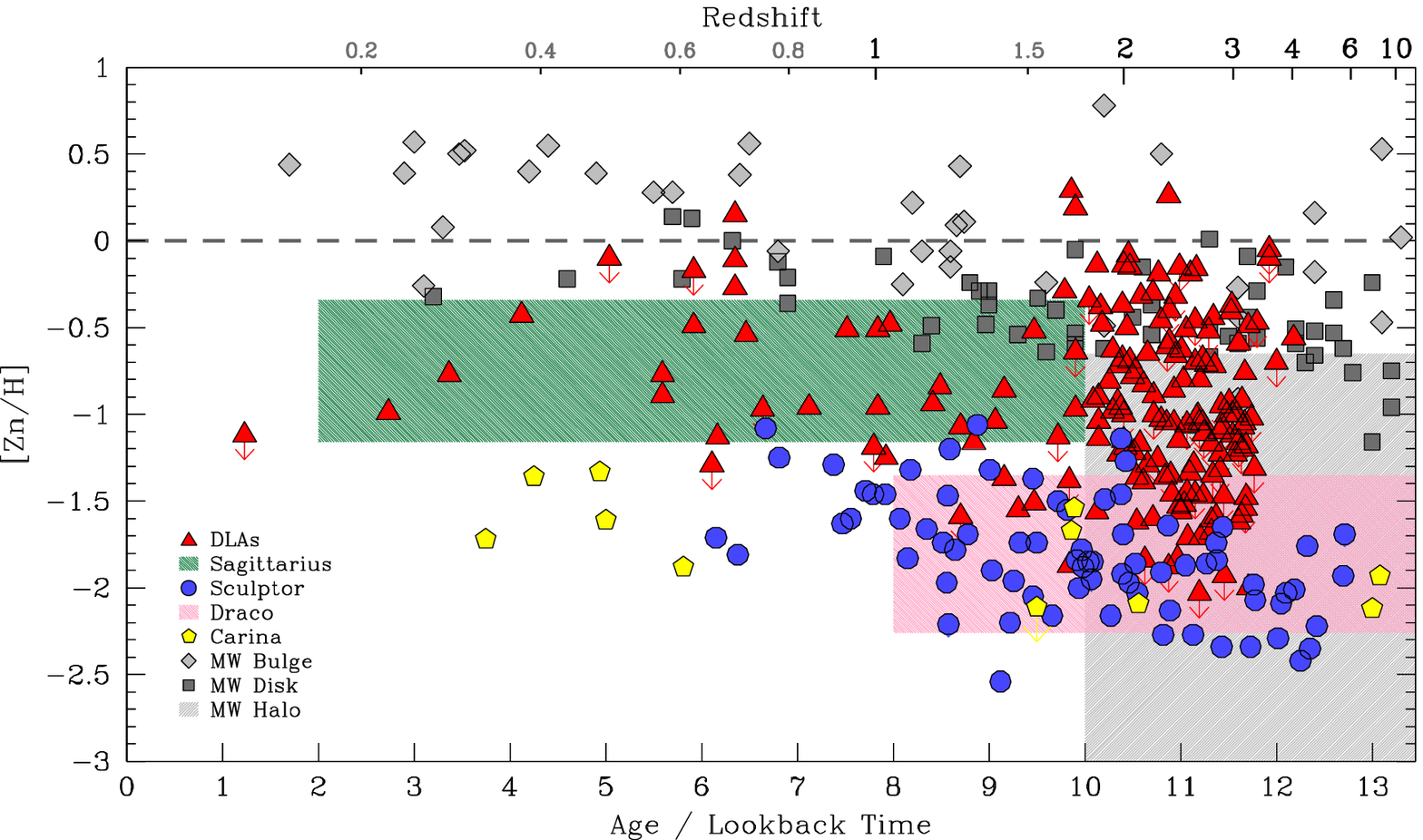}
      \caption{Zinc abundances as a function of stellar ages/look-back time (lower x-axis), and redshift (upper). Red triangles are DLAs (compilation from \citealt{Quiret2016}), blue circles are Sculptor stars (\citealt{Skuladottir2017}; ages from \citealt{deBoer2012}), and yellow pentagons are Carina stars (\citealt{Shetrone2003,Venn2012}; ages from \citealt{Lemasle2012}). Colored regions show the area occupied by available measurements: Draco in pink \citep{Shetrone2001,Cohen2009}; and Sagittarius in dark green \citep{Sbordone2007}. The Milky Way is shown in gray: diamonds are the bulge \citep{Bensby2013}, squares are the disk \citep{Reddy2006} and the halo is shown with a shaded area (\citealt{Cayrel2004,Reddy2006,Nissen2007}). 
      }
         \label{fig:azzn}
   \end{figure*}

Therefore it would be useful to have a way of measuring [$\alpha$/Fe] in DLAs, preferably using volatile elements to avoid uncertain corrections to account for dust depletion. Out of these, only O and S are $\alpha$-elements. The O~I line at 1302~\AA , which is typically used for O~abundance measurements in DLAs, is very strong, and becomes too saturated for accurate measurements at $\text{[O/H]}\gtrsim-1.5$. At higher metallicities, S is often the only tracer of $\alpha$-elements in DLA systems. The volatile element Zn is the heaviest of the iron-group. In the Milky Way, [Zn/Fe] is fairly constant over a large range in metallicities, $-2\lesssim \text{[Fe/H]}\lesssim0$, which has prompted the use of Zn as a proxy for Fe in DLA systems (e.g., \citealt{Pettini1990,Cia2016}) and thus traditionally [S/Zn] has been used as a tracer for [$\alpha$/Fe].

\begin{table}
\begin{threeparttable}
\caption{Stellar and dynamical masses of dwarf galaxies discussed in this paper (\citealt{McConnachie2012}, more details therein). Also listed are the number of available measurements of [Zn/H] and [S/Zn] in the literature. Upper limits and stars with abundance errors >0.5~dex are not included.}
\label{tab:dwarfgalaxies}
\centering
\small
\begin{tabular}{l c c c c c}
\hline\hline
Galaxy                  &       $M_\star$        &      $M_\textsl{dyn}$ & $N_\star$ & $N_\star$ & Ref.  \\
                                        &  10$^6$ M$_\odot$ & 10$^6$ M$_\odot$  & [Zn/H] & [S/Zn] &\\
\hline          
Sagittarius     &       21              &       190     & 11    & 0             &       1\\
Fornax                  &       20              &       56      & 2             & 0               &        2 \\
Sextans                 &       0.44             &       25      & 5             & 0               &        3,4\\
Sculptor                &       2.3             &       14      & 96    & 78              &        2,5 \\
Leo I                   &       5.5             &       12      &       2               &       0               &  2\\
Draco                   &       0.29            &       11      & 9             & 0               &       3,6\\
Ursa Minor      & 0.29          &       9.5     & 21    & 0             &       3,7\\
Carina                  &       0.38            &       6.3     & 10    & 0               &       2,8\\
Bootes I                &       0.03            &       0.8     & 3             & 0               &       9\\

\hline
\end{tabular}

\begin{tablenotes}
\scriptsize
\item \text{[1]} \citealt{Sbordone2007} [2] \citealt{Shetrone2003} [3] \citealt{Shetrone2001} [4] \citealt{Honda2011} [5] \citealt{Geisler2005,Kirby2012,Simon2015,Skuladottir2015b,Skuladottir2015a},2017; Hill et al. in prep. [6]  \citealt{Cohen2009} [7] \citealt{Sadakane2004,Cohen2010,Ural2015} [8] \citealt{Venn2012} [9] \citealt{Gilmore2013}.

\end{tablenotes}

\end{threeparttable}
\end{table}

Measuring S and Zn in stars in Local Group dwarf galaxies poses a new set of challenges. At the distances of the Milky Way dwarf satellites it is only feasible to take the high-resolution spectra, necessary for accurate abundance determination, of the brightest red giant branch (RGB) stars. The Zn abundance measurements are usually made from a Zn~I line at 4810~\AA, where the flux is relatively low in these RGB stars. At low metallicities, typical for stars in dwarf galaxies, S is usually measured from the S~I triplet at 9213, 9228, and 9238~\AA, a region that is relatively poor in atomic lines and thus seldom included when optical spectra of stars are acquired. In addition, these lines are sensitive to departures from local thermodynamical equilibrium, so non-LTE effects need to be taken into account in the derivation of the S abundance \citep{Takeda2005,Korotin2008,Skuladottir2015b}.

The chemical properties of the oldest stellar populations in nearby dwarf galaxies offer a direct insight into the star formation processes at early times, comparable to the redshift where most DLAs are observed. Because of these challenges, however, a moderately large sample of S and Zn measurements in stars of a dwarf galaxy has only recently become available in the Sculptor dwarf spheroidal (dSph) galaxy (\citealt{Skuladottir2015b,Skuladottir2017}). Thus for the first time we are able to make a direct comparison of the [S/Zn] ratios between DLA systems and stars in dwarf galaxies, without using any other element as a proxy.

\interfootnotelinepenalty=10000

\section{Metallicity comparison}

The available [Zn/H] abundance measurements of stars in the Local Group and gas in DLAs are shown in Fig.~\ref{fig:azzn} as a function of age/look-back time and redshift.\footnote{The solar abundances throughout this paper are adopted from \citet{GrevesseSauval1998}, with the exception of $\log \epsilon(\text{S})_\odot = 7.16$ \citep{Caffau2011}.} The conversion between redshift and look-back time was made using \citet{Wright2006}\footnote{The default values of \citet{Wright2006} are assumed ($H_0=69.6, \Omega_M=0.286$, $\Omega_\textsl{vac}=0.714$), but adopting those determined by the Planck mission \citep{Planck2016} would make very little difference for our present purposes.}. The measurement errors on the DLA redshifts are negligible compared to the uncertainty of the individual stellar ages, $<\Delta\text{Age}>_\textsl{Scl}=<\Delta\text{Age}>_\textsl{Car}=1.8$~Gyr, but some systematic uncertainty in the conversion between redshift and age may remain. 

Measurements of Zn from four Milky Way dSph satellites are shown in Fig.~\ref{fig:azzn}. Out of these, the Carina dSph is the smallest and Sagittarius is the largest and most metal-rich; see Table~\ref{tab:dwarfgalaxies}. The individual stars with [Zn/H] measurements in Draco and in the main body of Sagittarius do not have age determinations. However, the stellar population in Draco is predominantly old (e.g., \citealt{Aparicio2001}). The star formation history of Sagittarius is more extended, and the available Zn measurements are in metal-rich stars, which are expected to be of young and/or intermediate age (\citealt{Bellazzini2006,Siegel2007,deBoer2015}). The range of age and [Zn/H] covered by various components of the Milky Way are also shown in gray in Fig.~\ref{fig:azzn}. In the case of the bulge and disk, the surveys included age measurements, but for the halo an age range of 10-14~Gyrs is assumed (e.g., \citealt{Salvadori2010,Carollo2016}).

The information in Fig.~\ref{fig:azzn} is limited to the available data, so it does not necessarily give a complete overview of the star formation histories of the dwarf galaxies in question.  In particular, the Sculptor dSph stellar sample is from the center of this galaxy which is younger and more metal-rich compared to other regions. From the star formation history it is estimated that over 80\% of Sculptor's stellar population was formed more than 10~Gyr ago \citep{deBoer2012}, while the data are spread over a larger range of ages, from 6~to 12~Gyr. For Sagittarius, [Zn/H] measurements are not available in the older and more metal-poor component. 

        \begin{figure*}
   \centering
   \includegraphics[width=\hsize-1.2cm]{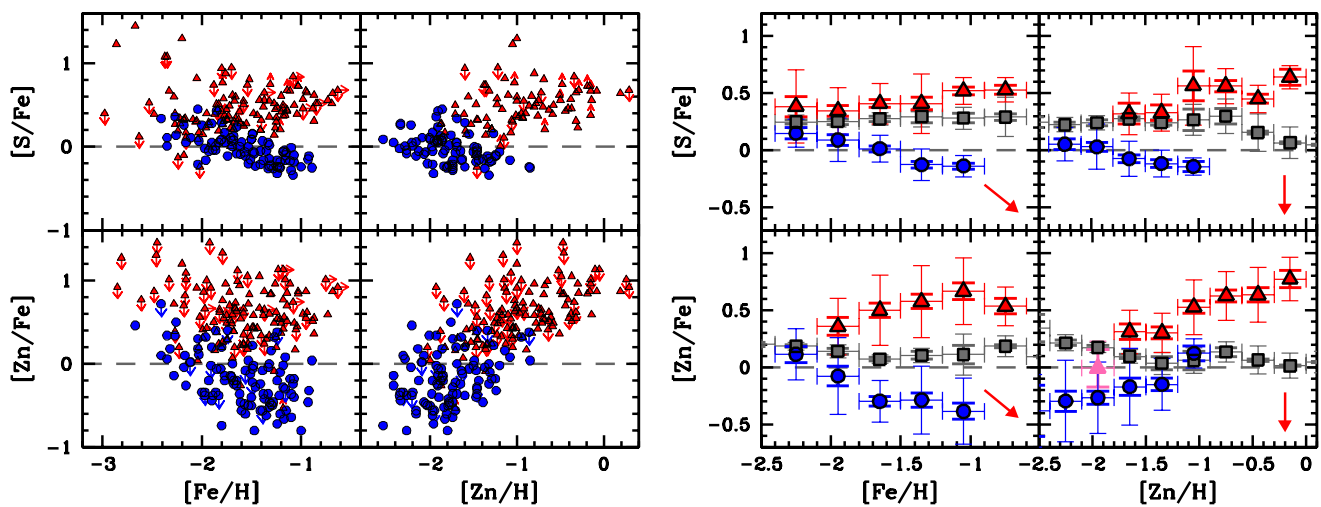}
      \caption{\textit{Left:} Abundance ratios of [S/Fe] and [Zn/Fe] in DLAs with $z<1.5$ (red triangles) and Sculptor stars (blue circles) as a function of [Fe/H] and [Zn/H]. \textit{Right:} Same with binned data, excluding upper/lower limits. Also included are Milky Way dwarf stars as gray squares. The size of each bin is shown with x-axis error bars. The thick y-axis error bars show the error of the mean in each bin, while thin show 1$\sigma$ of the scatter. The pink triangle only includes three DLAs. Red arrows show the direction (but not necessarily the size) of dust depletion corrections that should be applied to DLA abundances. We note that the range of both the x- and y-axis is not the same on the left and right panels. References: \textit{DLAs}: Compilation by \citealt{Quiret2016}; \textit{Sculptor:} \citealt{Skuladottir2017} (Zn), \citealt{Skuladottir2015b} (S), \citealt{Skuladottir2015a} (1 star); \textit{Milky Way:} \citealt{Reddy2003,Reddy2006}, \citealt{Nissen2007}.
      }
         \label{fig:nobinbin}
   \end{figure*}   

Most of the DLA data are concentrated in the redshift range $1.5<z<4$; see Fig.~\ref{fig:azzn}, which is when star formation in the Universe was at its peak \citep{Madau2014}. This corresponds to a look-back time of 9.4 to 12.2~Gyr, when star formation was active in all of the galaxies depicted in Fig.~\ref{fig:azzn}. At redshifts of $z\lesssim1.5$, and $z\gtrsim4$ [Zn/H] becomes very challenging to measure from the ground. 

The DLAs span a larger [Zn/H] range than any of the individual dwarf galaxies, as they are most probably a mixed bag of different types of systems, at different evolutionary stages \citep{Pettini2004,Pontzen2008,Fumagalli2015,Krogager2017}. The lower-metallicity DLAs, at $-2\lesssim\text{[Zn/H]} \lesssim-1$, are most compatible with the metallicities of dwarf galaxies such as Sculptor, Carina, and Draco, and the more metal-rich part of the Milky Way halo. The more metal-rich DLAs, $\text{[Zn/H]}\gtrsim-1$, fall in the same range as the Milky Way disk and bulge, and larger dwarf galaxies such as Sagittarius. At the lowest metallicities, $\text{[Zn/H]}\lesssim-2.5$, Zn becomes very challenging to measure, both in stars and DLAs. However, DLAs with undetectable Zn lines only amount to a small fraction of the population in this redshift range.

\section{Chemical abundance ratios with Fe} 
  
\subsection{[S/Fe] and [Zn/Fe]}

The available abundances of the key elements Fe, S, and Zn, measured in gas in DLAs and in RGB stars in the Sculptor dSph are shown in Fig.~\ref{fig:nobinbin}. The right panel displays the binned data, also including measurements of dwarf stars in the Milky Way. DLA systems are limited to $z>1.5$ because of observational difficulties at lower redshifts, and also because this corresponds to the epoch where the majority of Sculptor stars were formed ($\approx$90\% according to \citealt{deBoer2012}). When comparing these abundance ratios, it's important to remember that DLA abundances are measurements along one line of sight in the ISM of a (proto)galaxy, while Sculptor and Milky Way abundances come from individual stars. The binned data for DLAs therefore represent averages over many galaxies, while the stellar abundances are averages for stars within the same galaxy.

From these data it is evident that the Sculptor dSph, the Milky Way, and DLAs all show clear differences in their [S/Fe] and [Zn/Fe] ratios, both as a function of [Fe/H] and [Zn/H]. In addition, Sculptor stars cover a lower [Zn/H] range
compared to DLAs. This implies that stars in Sculptor 
are typically more metal-poor than gas in DLAs, even when observations are 
limited to the same age/redshift range (see also Fig.~\ref{fig:azzn}).
On the other hand, when looking at [Fe/H] abundances, we might 
have the misguided impression that DLAs cover the same metallicity range as 
stars in Sculptor. This is due to the fact that a fraction of Fe in DLAs is depleted onto dust and therefore not detected. Hence, it is extremely 
important to use volatile elements such as S and Zn when comparing 
metallicities in stars and gas.

The dust depletion of Fe also affects the [S/Fe] and [Zn/Fe] abundances, and the arrows in Fig.~\ref{fig:nobinbin} show the direction of dust depletion corrections that should be applied in each case. However, the size of these corrections is not well constrained but should be smaller at lower metallicities, because of less efficient dust formation (e.g., \citealt{Vladilo2002,Akerman2005,Jenkins2017}). Depending on the level of dust depletion, the DLA [S/Fe] or [Zn/Fe] abundance ratios could therefore resemble those of the Milky Way, or Sculptor, or neither.

                  \begin{figure}
   \centering
   \includegraphics[width=\hsize-0.5cm]{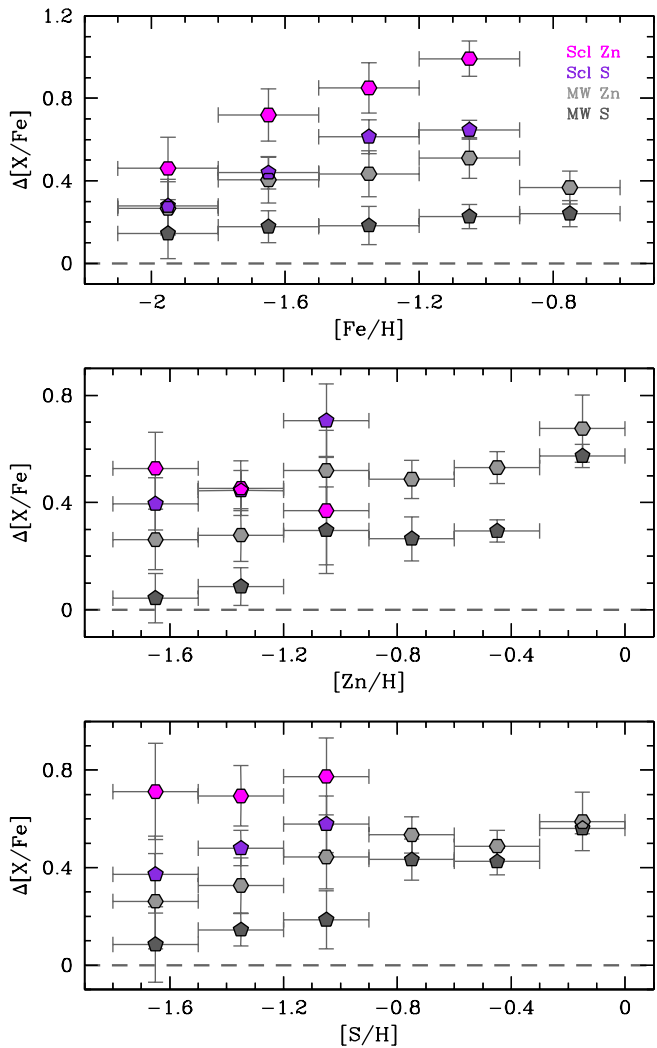}
      \caption{Differences between measured abundances in DLAs and stars as a function of [Fe/H] (top), [Zn/H] (middle) and [S/H] (bottom). Comparisons with Sculptor (Scl) stars are shown in pink (Zn) and purple (S), and with the Milky Way (MW) stars in light gray (Zn) and dark gray (S). Hexagons are $\Delta \text{[Zn/Fe]}$ and pentagons are $\Delta \text{[S/Fe]}$. Only DLAs and stars that have published measurements of all three elements, Fe, S, and Zn, are included. The y-axis error bars show the quadratic sum of the errors of the means for DLAs and Scl or MW, while x-axis error bars show the size of each bin. References: see Fig.~\ref{fig:nobinbin}.
      }
         \label{fig:delta}
   \end{figure}

\subsection{Can we quantify dust depletion?}

A standard approach to correct for dust depletion in DLAs is to assume that [Zn/Fe] ratios in DLAs are intrinsically the same as in the Milky Way stars at any given metallicity. Since S and Zn are not significantly depleted at the metallicities and H~I column densities of most DLAs \citep{Jenkins2017}, we can quantify the dust depletion correction as $\Delta\text{[Zn/Fe]}_\text{MW}=\text{[Zn/Fe]}_\text{DLA}-\text{[Zn/Fe]}_\text{MW}$ or equivalently by using [S/Fe] abundances. On the other hand, if it is assumed that [Zn/Fe] or [S/Fe] should behave the same in DLAs and the Sculptor dSph, then the dust depletion can be determined using $\Delta \text{[X/Fe]}_\text{Scl}$. 

The values of $\Delta \text{[X/Fe]}_\text{MW/Scl}$ are shown in Fig.~\ref{fig:delta}. It is clear that neither [Zn/Fe] nor [S/Fe] behave the same way in Milky Way and in Sculptor (see Fig.~\ref{fig:nobinbin} and \ref{fig:delta}), showing that these abundance ratios are not a constant throughout all environments, and casting doubt on whether this is a robust method for measuring dust depletion. 

Only DLAs and stars that have measurements of all  three elements, Fe, Zn and S, are shown in Fig.~\ref{fig:delta} as a function of [Fe/H], [Zn/H], and [S/H]. Out of these, the top panel is the most challenging to interpret, since we are not necessarily comparing the same metallicities in stars and DLAs, due to dust depletion of [Fe/H] in DLAs, so here we focus on the other two panels. In any given metallicity bin the average differences $\Delta\text{[Zn/Fe]}$ and $\Delta\text{[S/Fe]}$ refer to exactly the same set of DLA and stellar measurements. This means that even though different DLAs might suffer different levels of dust depletion, if [S/Fe] and [Zn/Fe] are both the same as in the Milky Way, the average dust depletion of the bin should be the same whether measured from $\Delta \text{[S/Fe]}$ or $\Delta\text{[Zn/Fe]}$, and as a function of either [S/H] or [Zn/H].

                  \begin{figure*}
   \centering
   \includegraphics[width=\hsize-2cm]{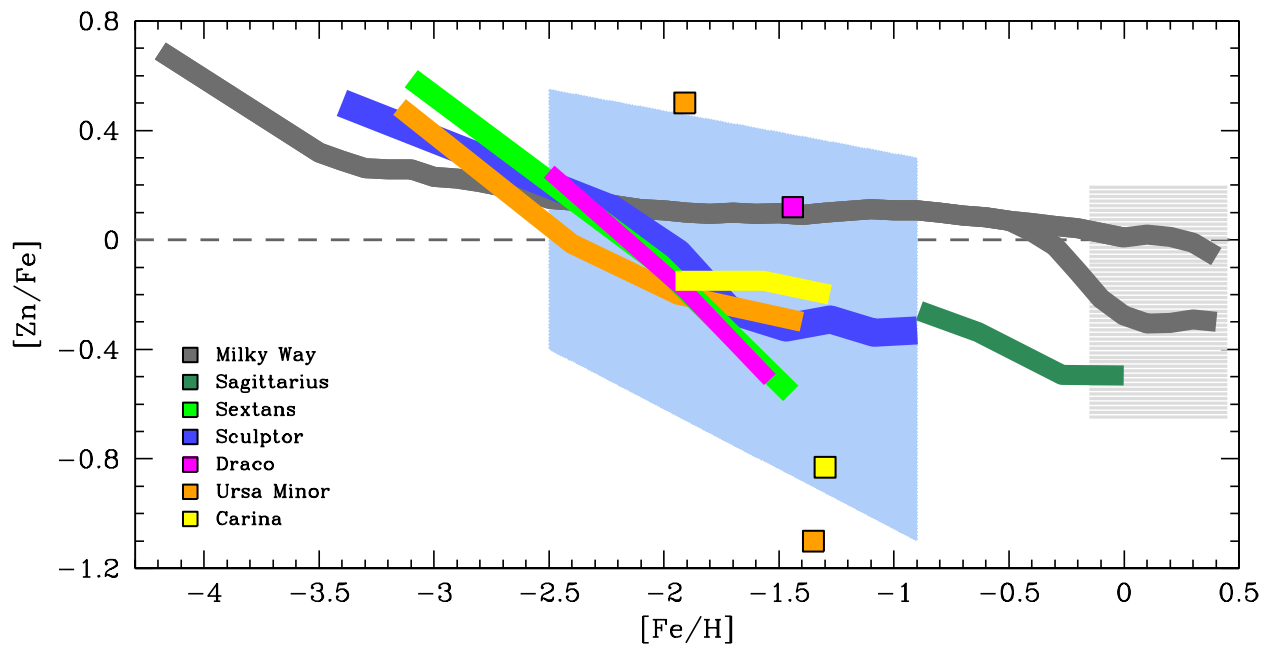}
      \caption{Schematic figure of [Zn/Fe] as a function of [Fe/H] in the Milky Way (gray) and its dwarf galaxy satellites: Sagittarius (dark green), Sextans (green), Sculptor (blue), Draco (pink), Ursa Minor (orange) and Carina (yellow). Lines show mean trends, while scatter is indicated by shaded areas, and notable outliers are identified by individual squares. The two branches at $\text{[Fe/H]}\gtrsim-0.5$ in the Milky Way represent the mean trends as measured by dwarf stars (higher values) and giant stars (lower values). Ref: \textit{Dwarf galaxies:} See Table~\ref{tab:dwarfgalaxies}. \textit{Milky Way:} \citealt{Reddy2003,Reddy2006,Cayrel2004,NissenSchuster2011,Ishigaki2013,Bensby2014,Barbuy2015,Duffau2017}.
      }
         \label{fig:lgzn}
   \end{figure*}

In the lowest panel of Fig.~\ref{fig:delta}, when [S/H] is used as a reference element, the values of dust depletion as measured from $\Delta \text{[S/Fe]}_\text{MW}$ and $\Delta \text{[Zn/Fe]}_\text{MW}$ agree within error bars, although the latter is always systematically higher. However, when [Zn/H] is used as a reference element (middle panel Fig.~\ref{fig:delta}), this is not the case. This means that there is no dust correction that can be applied to make both the [S/Fe] and [Zn/Fe] ratios agree between the Milky Way (or Sculptor) and DLA systems at any given metallicity. This can result from some or all of the following effects: i)~The chemical enrichment history of the Milky Way (or Sculptor) and the average of DLAs in each metallicity bin is not the same; ii)~there is some dust depletion present of S and/or Zn in DLAs; iii)~the stellar abundances are inaccurate (e.g., due to incomplete stellar physics applied); iv)~the dust correction of Fe is inhomogeneous in each DLA, and correlated with S and/or Zn abundance in each region along the line of sight, creating this effect.

\subsection{The complications with Zn}

It is clear from both Figs.~\ref{fig:nobinbin} and~\ref{fig:delta} that [Zn/Fe] is not the same in the Sculptor dSph galaxy and the Milky Way at all metallicites. The question thus becomes one regarding whether or not Zn abundances in Sculptor are particularly unusual. A schematic view of the available measurements of stars in Local Group dwarf spheroidal galaxies is shown in Fig.~\ref{fig:lgzn}, along with the different Milky Way components. It is immediately evident that [Zn/Fe] is not constant at all metallicities in any of these galaxies, including the Milky Way. In addition, dwarf galaxies typically exhibit different values of [Zn/Fe] at a given [Fe/H] compared to the Milky Way. All galaxies (with the possible exception of Carina) seem to show a trend of declining [Zn/Fe] with increasing [Fe/H]. For Sculptor, Draco, Sextans, and Ursa Minor this happens at a similar metallicity, while it is not completely clear where this trend lies in Sagittarus, due to limited data. With the exception of Sculptor, the mean trend in these dwarf galaxies is only determined by $5-21$ stars (see Table~\ref{tab:dwarfgalaxies}), and should thus only be interpreted as an indication. In particular, the Zn abundances cannot always be measured at $\text{[Fe/H]}<-2.5$. Thus, the rise in [Zn/Fe] at the lowest values of [Fe/H] in dwarf galaxies is highly uncertain, and may well be less pronounced than it appears in Fig.~\ref{fig:lgzn}.

To complicate matters even further, there is significant scatter in [Zn/Fe] at a given [Fe/H] in the Local Group dwarf spheroidal galaxies, which is not well understood; see Fig.~\ref{fig:lgzn}. The measurement of Zn in dwarf galaxies is challenging, and can only been done in giant stars,  and usually only one line is used. Therefore, some of the scatter in the Sculptor dSph probably reflects measurement uncertainties; however not all stars at a given [Fe/H] are consistent with having the same [Zn/Fe] while that is true for most [$\alpha$/Fe] measurements in this galaxy (e.g., \citealt{Skuladottir2015b,Skuladottir2017}, Hill et al. in prep). Outliers in [Zn/Fe] can also be found in other dwarf galaxies (see Fig.~\ref{fig:lgzn}), although data is limited.

At the lowest metallicities in the Milky Way, there is an increasing trend of [Zn/Fe] with decreasing [Fe/H]. At the highest metallicities, $\text{[Fe/H]}\geq-0.3,$ there is a discrepancy between [Zn/Fe] as measured in dwarf and giant stars in the Milky Way, where the latter show lower values \citep{Bensby2014,Barbuy2015,Duffau2017}. There is currently no consensus as to the reason for this discrepancy. One possibility is that it may be due to inadequacies in synthesizing line formation in stellar atmospheres.\footnote{All available calculations for the effects of non-LTE, 3D and granulations are unable to explain this difference, see \citet{Duffau2017}.} Perhaps most likely is that dwarf and giant stars at high metallicities probe different populations, as the giant stars are brighter and thus can be observed over greater distances (for a more detailed discussion, see \citealt{Duffau2017}). In the metallicity range $-2.5 \lesssim \text{[Fe/H]}\lesssim-0.5$, measurements of [Zn/Fe] in dwarf and giant stars in the Milky Way are in very good agreement, meaning that if the discrepancy arises from differences in stellar physics between dwarf and giant stars, these effects are only important at higher metallicities. At the lowest $\text{[Fe/H]}<-2.5$, Zn abundances have currently only been measured in giant stars.

Turning to theory, Zn is believed to be mainly produced in type II SN and high-energy SN, referred to as hypernovae \citep{HegerWoosley2002,Kobayashi2006}. Other production sites have also been proposed, such as the s-process \citep{Travaglio2004} and neutrino-rich wind of a newly formed neutron star \citep{Pruet2005}, or electron-capture SN \citep{Wanajo2018,Hirai2018}. The ejecta of type Ia SN, on the other hand, are thought to contain relatively small amounts of Zn, with abundance ratios as low as $\text{[Zn/Fe]}\lesssim-1.2$, accompanied also by subsolar ratios of $\alpha$-elements, [$\alpha$/Fe]<0 \citep{Iwamoto1999,Kobayashi2006}. These type Ia SN start to effect the chemical composition of their environment 1-2 Gyr after onset of star formation and thus lower [$\alpha$/Fe] and [Zn/Fe] in their surroundings. Since star formation in dwarf galaxies is less efficient than in the Milky Way, this happens at lower [Fe/H]. The decrease of [Zn/Fe], shown in Fig.~\ref{fig:lgzn}, corresponds to the same trends in [$\alpha$/Fe] in each galaxy (\citealt{Tolstoy2009} and references therein). Calculations using theoretical yields have also shown that the measured decrease of [Zn/Fe] and [$\alpha$/Fe] with [Fe/H] in the Sculptor dSph galaxy are consistent with being caused by a type Ia SN contribution to the abundance of  Fe \citep{Skuladottir2017}. 

In the Milky Way, where stars span a larger range of [Fe/H], the evolution of [Zn/Fe] is complicated by the fact that the yields of Zn in type II SN and hypernovae are very metallicity dependent, with a sharp increase in [Zn/Fe] at $\text{[Fe/H]}\gtrsim-1$ \citep{Kobayashi2006}, which counteracts the influence of type Ia SN at a similar metallicity in the Milky Way. This creates the illusion that Zn and Fe are co-produced, whereas the fairly constant value of [Zn/Fe] at $-2.5<\text{[Fe/H]}<0$ is the result of the particular star formation history of the Milky Way, and need not apply to other galaxies. The effects of type Ia SN on [Zn/Fe] can still be seen in the Milky Way by looking at stars with similar [Fe/H] but different [$\alpha$/Fe]. In these cases [Zn/Fe] typically correlates with [$\alpha$/Fe] (\citealt{NissenSchuster2011,Mikolaitis2017}).

   %
\section{Un-depleted chemical elements}

\subsection{Zn as a reference element}

\begin{figure}
   \centering
   \includegraphics[width=\hsize-0.8cm]{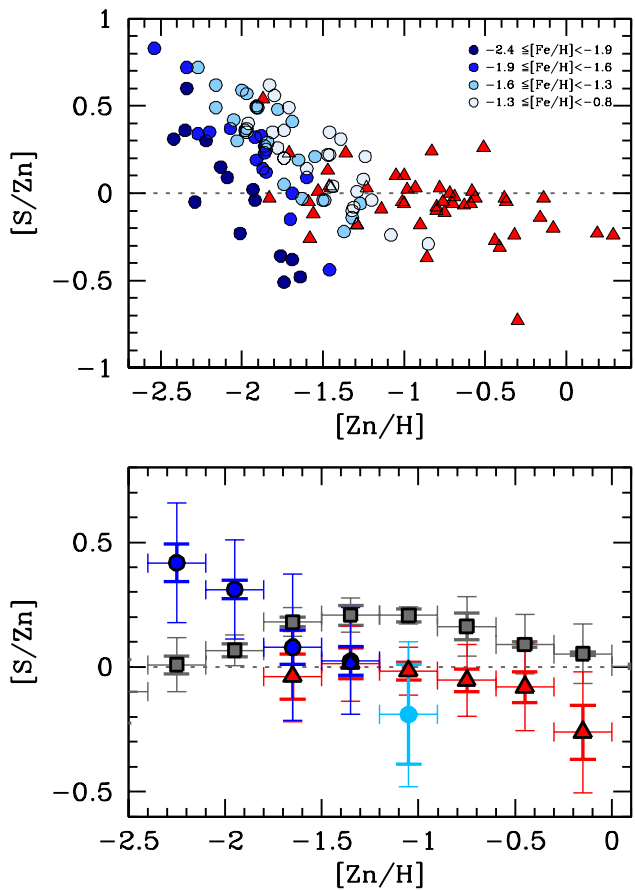}
      \caption{Abundance ratios of [S/Zn] as a function of [Zn/H] for DLAs ($z>1.5$), as well as Sculptor and Milky Way stars. In the top panel, Sculptor stars are circles, colored according to their ages: the oldest stars are dark blue, and the youngest are light blue. Stars without age determination are represented by unfilled circles. The light blue circle in the bottom panel only includes three Sculptor stars. Otherwise, symbols are the same as in Fig.~\ref{fig:nobinbin}, where references are also listed.
      }
         \label{fig:SZn}
   \end{figure}

The abundance ratios of [S/Zn] as a function of [Zn/H] are shown for DLAs, the Sculptor dSph, and the Milky Way in Fig.~\ref{fig:SZn}. As previously noted, Sculptor stars cover a different [Zn/H] range compared to the DLAs. However, where these samples overlap, their [S/Zn] abundance ratios are in good agreement, while the Milky Way shows a slightly different trend. A straightforward interpretation could be that the lower metallicity DLAs correspond to systems very similar to the Sculptor dwarf spheroidal galaxy. 

However, the complexity of the evolution of Zn in Local Group galaxies (shown in Fig.~\ref{fig:lgzn} and discussed in Section~3.3) leads us to question the value of [Zn/H] as a reference for metallicity (at least in dwarf galaxies). Ideally, an element used as a proxy for metallicity should be: i) reliably measured; ii) strongly connected to the evolutionary state of the galaxy or its component; iii) of well understood nucleosynthesis. From the point of view of stars in dwarf galaxies, [Zn/H] fails on all fronts, as it is difficult to measure, thus having large abundance measurement errors. In addition, there is an intrinsic scatter, which is often difficult to distinguish from uncertainties, and the theoretical understanding of Zn production in dwarf galaxies is very limited. 

Since Zn does not trace type~Ia SN (see Section~3.3), and has a large scatter for a given [Fe/H] (both intrinsic and due to measurement errors), [Zn/H] abundances have a relatively weak correlation to the stellar ages in Sculptor, as shown with color coding in Fig.~\ref{fig:lgzn}. Thus, any abundance ratios in Sculptor (and other dwarf galaxies) become very difficult to interpret as a function of [Zn/H].

                  \begin{figure*}
   \centering
   \includegraphics[width=\hsize-3.cm]{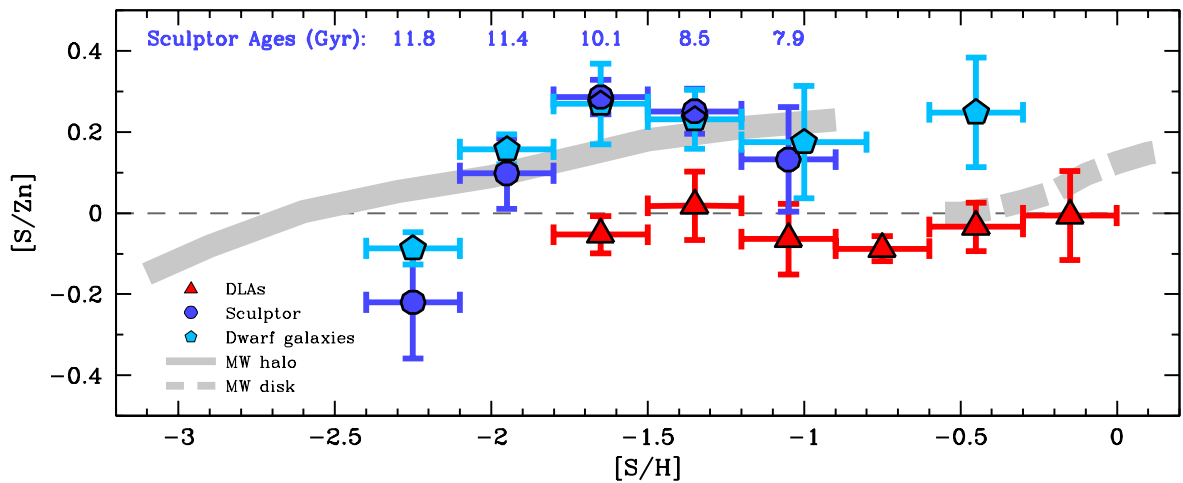}
      \caption{Abundance ratios of S and Zn for DLAs (red triangles), Sculptor stars (blue circles) and other local dwarf galaxy stars (light blue pentagons). Average ages of Sculptor stars are listed at the top (blue) for each bin \citep{deBoer2012}. Error bars show the size of the bin (x) and the error of the mean (y). Mean trends of Milky Way stars are shown with gray lines: solid for the halo (dwarf and giant stars); dashed for the disk (dwarf stars).  With the exception of Sculptor, a proxy for S is adopted for all dwarf galaxy stars: $\text{[S/H]}=\frac{1}{2}(\text{[Ca/H]}+\text{[Ti/H]}$). 
      References: \textit{DLAs:} \citealt{Quiret2016}; \textit{Dwarf galaxies:} See Table~\ref{tab:dwarfgalaxies}; \textit{Milky Way halo:} \citealt{Cayrel2004,Nissen2004,Nissen2007,Spite2011}; \textit{Milky Way disk:} \citealt{Reddy2003}.
      }
         \label{fig:SZnall}
   \end{figure*}

\subsection{S as a reference element}

As Zn is a problematic reference element, we propose the use of [S/H] instead. Although S is an $\alpha$-element, a non-negligible amount is created in type Ia SN. Therefore, S traces both type Ia and  type II SN, similar to Fe, but in different proportions. The exact fraction of the contribution from each SN type depends on the star formation history of the system and the adopted models and yields, but as an example, \citet{Tsujimoto1995} predicted around 25\% of S in the solar neighborhood to be created by type~Ia SN, while the fraction for Fe is around 60\%. Therefore, although S should not be used interchangeably with Fe, these two elements trace the same processes. 

Compared to other elements, however, S is relatively difficult to measure in DLAs, as the available S~II lines (1259, 1253, and 1250~\AA) are weak and in many cases fall within the Lyman-$\alpha$ forest, where blending can be a problem. Other choices, in particular Si, might also be a good reference element for DLAs, but the available Si measurements for stars in the Sculptor dSph are very limited (see Appendix A), and therefore not suitable for our present purpose. In addition, Si is predicted to suffer more dust depletion compared to S (e.g., \citealt{Cia2016}).

The relation between Zn and S as a function of [S/H] is shown in Fig.~\ref{fig:SZnall} for DLAs, the Milky Way, Sculptor and other dwarf galaxies. Since measurements of these elements in other dwarf galaxies are scarce, they are binned together in Fig.~\ref{fig:SZnall} using a proxy: $\text{[S/H]}\approx\frac{1}{2}(\text{[Ca/H]}+\text{[Ti/H]})$ (see Appendix A), as there are no available S measurements for these stars. As shown by the average Sculptor stellar ages \citep{deBoer2012} listed in Fig.~\ref{fig:SZnall}, [S/H] is a reasonable tracer of the evolution of this galaxy. The age-[S/H] relation in the other dwarf galaxy sample is less clear, however, as it contains the averages of many different galaxies. In the case of DLAs, there is not a clear age/redshift trend with [S/H], as there is a very large metallicity scatter at any redshift (e.g., \citealt{Rafelski2012,Rafelski2014}), but the DLA sample in Fig.~\ref{fig:SZnall} is limited to $z>1.5$ ($\text{age}>9.4$~Gyr).

The trend of increasing [S/Zn] with [S/H] in Sculptor stars can be interpreted as resulting from an added contribution to the S abundance by type~Ia SN which produce very small amounts of Zn compared to S \citep{Iwamoto1999,Kobayashi2006}. For a more detailed discussion and calculations of this, see \citet{Skuladottir2017}. The ratio [S/Zn] appears to `saturate’ at \linebreak $\approx$~$+0.2$ dex at $\text{[S/H]}>-1.5$, that is, for stars with ages $<10$~Gyr. The reason for this behavior is not well understood; possibly it may be caused by a metallicity dependence of Zn yields in type II SN, or other unidentified processes adding to the Zn production, such as neutrino-driven winds \citep{Hoffman1996} or the weak and/or main s-processes \citep{Mishenina2002,Travaglio2004}.

The behavior of [S/Zn] versus [S/H] seen in the other Local Group dwarf galaxies considered here seems on average to be similar to that seen in Sculptor (see Fig.~\ref{fig:SZnall}). However, we note that this result might be biased, due to lack of data. With the exception of the most [S/H]-rich bin, the data shown in Fig.~\ref{fig:SZnall} are dominated by dwarf galaxies of similar size or smaller than the Sculptor dSph (see Table~\ref{tab:dwarfgalaxies}). If Zn behaves similarly to $\alpha$-elements in the larger dwarf galaxies, Sagittarius and Fornax (as Zn does in the smaller ones), subsolar [S/Zn] are expected at higher [S/H] compared to Sculptor and the sample shown in Fig.~\ref{fig:SZnall}.

Turning to the Milky Way, we see in Fig.~\ref{fig:SZnall}. a very complex structure in [S/Zn] versus [S/H]. In halo stars, [S/Zn] decreases with decreasing [S/H], which has been interpreted as evidence for Zn production in hypernovae, which can produce large amounts of Zn \citep{UmedaNomoto2002}.  Disk stars seem to be disjointed from this behavior in the halo, and more similar to DLAs. The lower values of [S/Zn] when $\text{[S/H]} > -1$ in the Milky Way disk have been explained by more efficient Zn production in type II SN and hypernovae at higher metallicities \citep{Kobayashi2006,Nomoto2013}.

Given the complexity of the [S/Zn] ratio in different Local Group stellar populations, it is remarkable to see in Fig.~\ref{fig:SZnall} an essentially constant mean value at all the metallicities probed by DLAs. In particular since the DLAs here are expected to cover a wide range in mass. By adopting the mass-metallicity relation of \citet{Christensen2014} and assuming an average redshift $z=2.4$ for the sample here, DLAs around $\text{[M/H]}\approx-0.1$ are expected to have a typical stellar mass of $\sim10^{11}$ M$_\odot$, while at $\text{[M/H]}\approx-1.2$, this is $\sim10^9$~M$_\odot$. Smaller, Sculptor-like DLAs with a stellar mass of $\sim10^{6}$ M$_\odot$ should thus presumably be found at lower metallicities, $\text{[S/H]} \leq - 2$, where data are currently lacking. A continuation of the constant [S/Zn] trend would imply that the most metal-poor DLAs are consistent with the early stages of dwarf galaxies like Sculptor.

Comparison with DLAs is further complicated by the fact that, in general, galaxies are not detectable as DLAs through their entire evolutionary history. For example, \citet{SalvadoriFerrara2012} predict that most Sculptor-like galaxies have typically already lost too much gas at $z=2.3$ ($\text{age}=10.8$~Gyr) to be detectable as DLAs. In Fig.~\ref{fig:SZnall} this corresponds to $\text{[S/H]}>-1.8$, where [S/Zn] abundance ratios are significantly higher in Sculptor than in DLAs. It is therefore plausible that some DLA galaxies may have undergone a chemical evolution similar to what we see in Sculptor, but are no longer detectable as DLAs by the time we observe them (and are therefore missing from DLA samples). How long a given galaxy retains its H~I gas and is detectable as a DLA highly depends on its environment and star formation history.

\subsection{An alternative to [$\alpha$/Fe]}

As a measurement of relative type Ia and type II SN contributions, we propose the use of the [O/S] abundance ratio. Both these elements are volatile, and as some S is produced in type Ia SN and O is essentially not, we expect the behavior of [O/S] with [S/H] to show qualitative (but not necessarily quantitative) similarities with the more often considered trend of [$\alpha$/Fe] with [Fe/H].

                 \begin{figure}
 \includegraphics[width=\hsize-0.5cm]{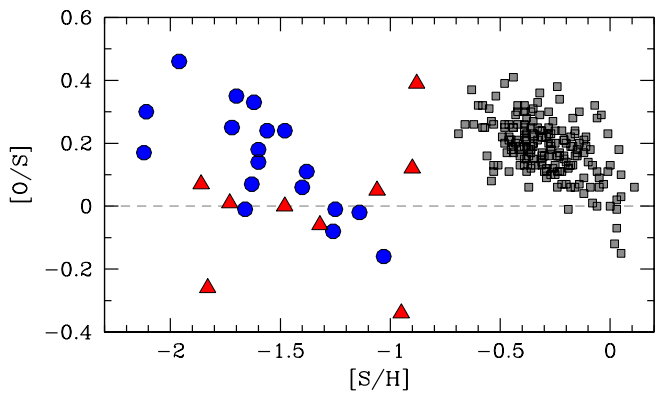}
      \caption{The abundance ratio of [O/S] as a function of [S/H] for DLAs (red triangles), Sculptor stars (blue circles) and Milky Way stars (gray squares). References: \textit{DLAs:} \citealt{Quiret2016}. \textit{Sculptor:} \citealt{Skuladottir2015b}, Hill et al. in prep. \textit{Milky Way:} \citealt{Reddy2003}.
     }
       \label{fig:OS}
   \end{figure}

The available data for DLA systems, Sculptor, and the Milky Way disk are shown in Fig.~\ref{fig:OS}. As predicted, the trend of [O/S] with [S/H] mimics that of [$\alpha$/Fe] in Sculptor. The same is true for the Milky Way where the `turnover’ to a trend of decreasing [O/S] with increasing [S/H] occurs at higher values than in the dwarf galaxies (as is the case for [$\alpha$/Fe] versus [Fe/H]). In the metal-poor halo, these values are generally supersolar with $<\text{[O/S]}>=0.46\pm0.14$ \citep{Cayrel2004,Spite2011}.

High [O/S] abundance ratios in DLAs should therefore correspond to systems that have not experienced a significant contribution by type Ia SN, while those with low [O/S] have. Since the timescales of type Ia SN are $\approx 1-2$~Gyr, DLAs observed at redshifts $z>3$ should have supersolar [O/S] ratios, and the same is expected at $\text{[S/H]}<-2$, from what is observed in the Milky Way halo and its satellites. Unfortunately the available measurements of [O/S] in DLA systems are very limited, mainly since O is challenging to measure at high metallicities, and S at low. Furthermore these elements are typically measured from different ionization states, O~I and S~II lines, possibly causing further complications.

Theoretical yields (e.g., \citealt{Tsujimoto1995})  predict a similar behaviour of [O/Si] versus [Si/H], and this is supported with Milky Way data (see Appendix). However [O/Si] is expected to be slightly less sensitive to type Ia SN contributions compared to [O/S]. With the use of Si also comes the added complication of dust depletion in DLAs. However, in the sample of metal-poor DLAs (where dust depletion is less effective)  from \citet{Cooke2011}, supersolar values are observed: $<\text{[O/Si]}>=0.11\pm0.06$ at $-3<\text{[Si/H]}<-2$. This is consistent with only moderate type Ia SN contribution. In comparison, the metal-poor Milky Way halo has $<\text{[O/Si]}>=0.30\pm0.11$ at similar metallicities \citep{Cayrel2004}.

\section{Consequences for previous studies}

Other studies have previously tried to use the volatile elements S and Zn to make comparisons between the chemical properties of DLAs and stars in the Local Group (e.g., \citealt{Nissen2007,Spite2011}). In particular, the link between DLAs and local dwarf galaxies was explored in detail in recent work by \citet{Cooke2015} and \citet{Berg2015}. However, the chemical abundance comparison in these studies was somewhat limited by lack of data in Local Group dwarf galaxies, as very few [Zn/H] and no [S/Zn] measurements were available. Both \citet{Berg2015} and \citet{Cooke2015} therefore used [Zn/H] and [Fe/H] interchangeably.

\begin{figure}
 \includegraphics[width=\hsize-0.5cm]{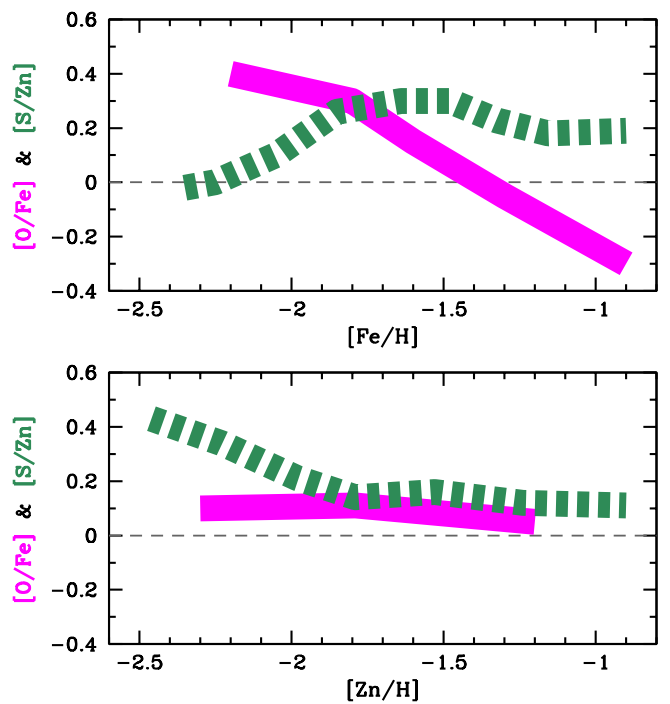}
      \caption{Schematic view of the mean trends of [O/Fe] (magenta solid lines) and [S/Zn] (green dashed lines) in the Sculptor dSph as a function of [Fe/H] (top) and [Zn/H] (bottom). \textit{References:} [O/Fe]: Hill et al. in prep. [S/Zn]: \citealt{Skuladottir2015b}, 2017.
     }
       \label{fig:BC}
   \end{figure}

The main new insight provided by the [S/Zn] data in Sculptor is the complicated nature of Zn abundances in dwarf galaxies, as captured in Fig.~\ref{fig:lgzn}. Using [Zn/H] as a proxy for [Fe/H] therefore leads to significant differences in chemical patterns. To illustrate this effect, schematic views of the mean trends of [O/Fe], a typical [$\alpha$/Fe] tracer, and [S/Zn] abundance ratios in Sculptor are shown both as a function of [Fe/H] and [Zn/H] in Fig.~\ref{fig:BC}. These abundance ratios show significantly different trends, both from each other, and whether [Fe/H] or [Zn/H] is used as a reference element. In conclusion: [S/Zn] does not show a similar relation with metallicity as [O/Fe], or other [$\alpha$/Fe] ratios do. Relevant discussion in previous comparison studies should therefore be revisited.

In particular, \citet{Cooke2015} showed [O/Fe] as a function of [Fe/H] for $\text{[Fe/H]}\lesssim-2$, and [S/Zn] as a function of [Zn/H] for $\text{[Fe/H]}\gtrsim-2$, assuming that both abundance ratios were tracing [$\alpha$/Fe] versus [Fe/H] in DLAs. By doing this, they observed a `knee' in these abundance ratios where they started to decrease at $\text{[Fe/H]}\approx-2$, analogous to that observed in [$\alpha$/Fe] as a function of [Fe/H] in dwarf galaxies. However, it is clear from Fig.~\ref{fig:BC} that this assumption does not hold (see also Section 4). Therefore, although a `knee'  in [$\alpha$/Fe] abundance ratios in DLAs cannot be excluded at this point, further confirmation is needed, preferably using the same probe throughout the entire metallicity range.

\section{Conclusions}

For the first time we have made a detailed chemical comparison between DLAs and stars in dwarf galaxies by using the volatile elements S and Zn without the help of any other element as proxy. This was made possible by recent measurements of S and Zn abundances in around 80 RGB stars in the Sculptor dwarf spheroidal galaxy \citep{Skuladottir2015b,Skuladottir2017}. Most of the stars in this galaxy formed $>10$~Gyr ago, which correspond to the redshifts at which the majority of DLAs are observed (i.e., $z\gtrsim1.8$).

Our analysis of the Sculptor data shows that S abundances behave as expected of an $\alpha$-element, primarily created by core-collapse SN. However, there is also a non-negligible contribution to S from the yields of type Ia SN. This makes [S/H] a reasonable tracer of the metallicity evolution in galaxies. In low-metallicity stars, S measurements are challenging, so when not available, other $\alpha$-elements can be used as proxy, preferably: Ca, Ti or the average of the two (see Appendix).
The lighter $\alpha$-elements, O and Mg, should not be used as proxies for S since, unlike S, they have negligible contribution from type Ia SN.

In light of the available measurements in the Milky Way and its satellite dwarf galaxies, it is clear that [Zn/Fe] is neither the same at a given metallicity in all galaxies probed here, nor constant in a given galaxy throughout its entire evolutionary history. The nucleosynthetic  channels that lead to the formation of Zn remain unclear, but the following statement is in agreement with both theory and observations: 
\textit{ Type Ia SN are not significant producers of Zn.} The formation channels of Zn are therefore not the same as Fe, and the [Zn/Fe] ratio is highly dependent on the star formation history of each system. Thus, using [Zn/H] as a proxy for [Fe/H] is fraught with potential pitfalls.

Furthermore, [S/Zn] should not be used as a proxy for [$\alpha$/Fe], which is commonly interpreted as indicative of the relative contributions from core-collapse (such as type II SN) and type Ia SN. Instead we suggest [O/S] with [S/H], which shows qualitatively the same trend as [$\alpha$/Fe] with [Fe/H] both in the Sculptor dwarf spheroidal galaxy and in the Milky Way. Both O and S are volatile elements that are not expected to be significantly depleted onto dust, and O traces mainly massive stars exploding as type II SN, while S is formed both in type Ia and II SN. We therefore predict supersolar [O/S] in DLAs at low metallicities, $\text{[S/H]}<-2$, and in particular at $z>3$, as type Ia SN have not yet had time to significantly enrich the environment.

Assuming that dust depletion of S and Zn in DLAs is negligible, we conclude that DLAs on average do not have both the same [S/Fe] and [Zn/Fe] as a function of [Zn/H] as either the Milky Way or the Sculptor dwarf spheroidal galaxy. Furthermore, over the full range of [S/H] metallicities considered here, $-2 \lesssim \text{[S/H]} \lesssim 0$, the [S/Zn] relative abundances of DLAs do not resemble the patterns found in either the Milky Way nor its dwarf galaxy satellites. Thus, DLA systems on average do not have the same star formation history at all metallicities, as the Milky Way nor Sculptor, nor any other dwarf galaxy where data are sufficient. This should not be surprising, as DLA systems are likely a mixed bag of objects, possibly with different properties at different metallicities. Therefore there is no reason to expect their star formation history to always follow that of the Milky Way or any other galaxy. 

Considering the different formation scenarios of S and Zn, and the complexity observed in stars in the Local Group galaxies, it is remarkable that the [S/Zn] ratio is on average constant at approximately the solar value over the full metallicity range probed by DLAs, $-1.8\lesssim\text{[S/H]}\lesssim0$ (see Fig. \ref{fig:SZnall}). It is possible that this is simply the result of averaging over galaxies with different star formation histories in each metallicity bin, or that we are looking at different galaxies at a similar stage in their evolution. Until we gather more information about the host galaxies of DLAs and the nucleosynthesis of Zn in particular, the lack of any trend in [S/Zn] versus [S/H] in DLAs remains a puzzle.

Finally, we note that although Zn is not interchangeable with Fe, it can still be reliably used as a metallicity indicator of DLA systems. The lowest relative [Zn/Fe] abundances we have measured in the Local Group are in galaxies that have lost their gas, and the highest [Zn/Fe] are measured at very low [Fe/H], where Zn is very challenging to measure in gas (see Fig. \ref{fig:lgzn}). It is therefore unlikely that these extremes are very common in DLAs. Away from such extremes, Zn should be a reasonably good metallicity indicator.

\begin{acknowledgements}
The authors are indebted to the International Space Science Institute (ISSI), Bern, Switzerland, for supporting and funding the international team `First stars in dwarf galaxies'. \'A.S. thanks S. Quiret for providing his DLA literature compilation, R. Cooke for insightful suggestions, and P. Noterdaeme for useful help. \'A.S. acknowledges funds from the Alexander von Humboldt Foundation in the framework of the Sofja Kovalevskaja Award endowed by the Federal Ministry of Education and Research. E.T. gratefully acknowledges support to visit to the IoA from the Sackler Fund for Astronomy. S.S. was supported by the European Commission through a Marie-Curie Fellowship, project PRIMORDIAL, grant nr. 700907, and by the Italian Ministry of Education, University, and Research (MIUR) through a Rita Levi Montalcini Fellowship. M.P. is grateful to the Kapteyn Astronomical Institute for their generous hospitality during multiple visits while this work was in progress.
\end{acknowledgements}

\bibliography{heimildir}

\clearpage
\pagebreak

\appendix

   %
\section{Proxy for sulphur in stars} 
  
\begin{figure*}
   \centering
   \includegraphics[width=\hsize-2.5cm]{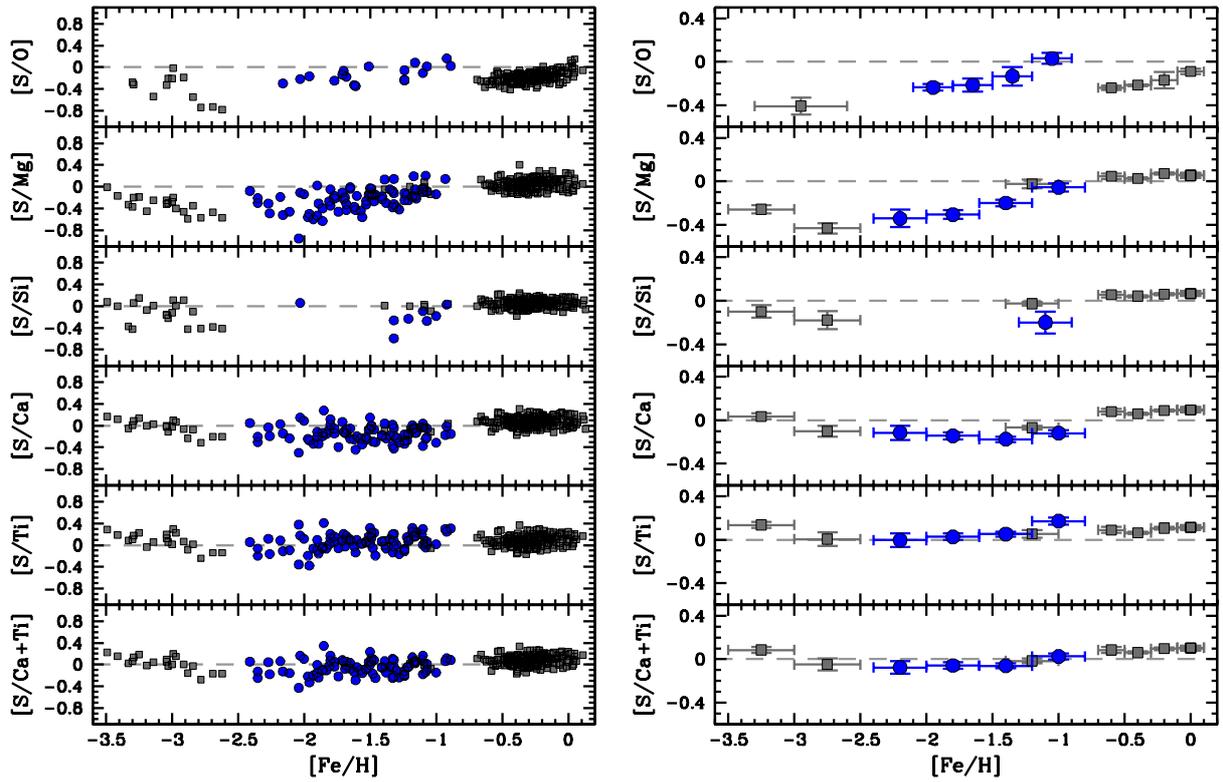}
      \caption{Stellar measurements in Sculptor (blue circles) and the Milky Way (gray squares) of abundance ratios of sulphur to other $\alpha$-elements: O, Mg, Si, Ca, Ti, and [S/(Ca+Ti)]= $\frac{1}{2}$([S/Ca]+[S/Ti]). Left panels show individual measurements while right panels show binned data, where the x-axis error bar shows the size of each bin and y-axis error bar is the error of mean. We note that the range on the y-axis is not the same in left and right panels. References:  \textit{Sculptor:} \citealt{Skuladottir2015a} (1 star), \citealt{Skuladottir2015b} (S), \citealt{Skuladottir2017} (Ti), Hill et al. in prep (Mg, Si, Ca). \textit{Milky Way:} \citealt{Chen2002,Reddy2003,ChenZhao2003,Cayrel2004,Nissen2007,NissenSchuster2010,Spite2011}.
      }
         \label{fig:salpha}
   \end{figure*}

The available S measurements in stars in dwarf galaxies are relatively scarce (see Table~\ref{tab:dwarfgalaxies}). Therefore, it is tempting to try to find another $\alpha$-element in stars that is suited for comparison with S in DLA systems. Comparisons of sulphur with other $\alpha$-element measurements in Sculptor and the Milky Way are shown in Fig.~\ref{fig:salpha}. Notably, not all [S/$\alpha$] abundance ratios show the same trends with [Fe/H], in either system. 
Larger deviations from a flat [S/$\alpha$] abundance ratio make the $\alpha$-element in question less suited to be used as a proxy for sulphur.

As already reported in \cite{Skuladottir2015b}, [S/Mg] in Sculptor shows an increasing trend with [Fe/H], and the same is true for [S/O]. This is consistent with theoretical yields which predict some production of S by type Ia SN, but negligible contribution to both O and Mg (e.g., \citealt{Tsujimoto1995,Iwamoto1999}).  Thus the ratios of [S/Mg] and [S/O] are higher in the yields of type Ia SN compared to  type II SN, so as the pollution by the former becomes more significant, the values of these abundance ratios increase. The trend in Sculptor of these ratios is therefore opposite to that of [$\alpha$/Fe] ratios observed at the same metallicity in this galaxy. Also in agreement with this explanation, [S/O] and [S/Mg] are not the same in the metal-poor halo and in Milky Way disk stars, confirming observationally that O and Mg are not ideal proxies for S. 

There are too few measurements of Si in Sculptor to draw any strong conclusions of how well it traces S. However, although there is only one data point at $\text{[Fe/H]}=-2$ this measurement is from higher-quality data (higher resolution and longer wavelength range) compared to the other Si measurement in Sculptor (\citealt{Skuladottir2015a}, Hill et al. in prep). By trusting this one data point, the trend with [Fe/H] seems weak, suggesting that Si is at least a better option than both O and Mg. The differences of [S/Si] at the metal-rich and metal-poor ends in the Milky Way are small, and seem to indicate that Si traces S reasonably well. This is consistent with theory. \citet{Tsujimoto1995} computed the relative contributions by type Ia and II SN for various elements and found that 25\% of S  and 17\% of Si in the solar neighborhood came from type Ia SN (while it was 1\% for both O and Mg). 

The elements Ca and Ti seem to trace S reasonably, both in Sculptor and the Milky Way, with $-0.2<\text{[S/Ca]}<0.1$ and $-0.1<\text{[S/Ti]}<0.2$. Both these elements have the advantage that they are relatively easy to measure in stars as there are many available lines in the optical part of the spectrum. There is a minor increasing trend in Sculptor of [S/Ti] with metallicity, while [S/Ca] is consistent with a flat, or a slightly decreasing trend. \citet{Tsujimoto1995} predicted that 25\% of Ca in the solar neighborhood originates in type Ia SN, which is exactly the same value as for S. Their study did not include Ti, but it has proven difficult  to calculate theoretical Ti yields that fit observational data (e.g., \citealt{Kobayashi2006}). However, observationally, both these elements seem to trace S reasonably well. Another possibility is to use the average of the two elements, and in all cases $-0.1<\text{[S/(Ca+Ti)]}$<0.1, where the ratio is defined as $\frac{1}{2}\text{([S/Ca]+[S/Ti])}$; see bottom panels of Fig.~\ref{fig:salpha}. Using two elements also has the advantage of reducing measurement errors.

It is important to keep in mind that each tracer has its own limitations and that Ca and Ti are not completely equivalent. In Sculptor, [Ca/Ti] shows a weak increasing trend with [Fe/H], indicating that the ratio [Ca/Ti] is not the same in type Ia  SN as in type~II SN, and/or that the metallicity dependence of the yields of these two elements is different. A similar result is seen in the Milky Way halo, where \cite{NissenSchuster2010} reported two distinct halo populations in the solar neighborhood, with different values of [$\alpha$/Fe]. Similar to Sculptor, the halo stars with low-$\alpha$ have a high value of $<\text{[Ca/Ti]}>_\text{low-$\alpha$}=0.098\pm0.006$, compared to the high-$\alpha$ population, $<\text{[Ca/Ti]}>_\text{high-$\alpha$}=0.042\pm0.005$. When comparing stellar and DLA abundances it is thus advised to consistently use the same proxy for S in all the stars involved, to decrease differences in systematic errors.

In conclusion, S in DLA systems should ideally be compared to S in stars. When this is not an option, using Ca, Ti, or the average of the two compared to the solar value should be the best alternative, judging from the available data in Sculptor and the Milky Way.

\end{document}